\documentclass[sigconf,nonacm]{acmart}
\setcopyright{none}
\acmConference{}{}{}
\acmDOI{}
\acmISBN{}

\usepackage{fancyvrb}
\usepackage{graphicx} % Required for inserting images
\usepackage{float}    % Add this package for forcing placement

\usepackage[percent]{overpic}
\usepackage{amsmath}
\usepackage{makecell}
\usepackage{algorithmic}
\usepackage{textcomp}
\usepackage{xcolor}
\usepackage{comment}
\usepackage{subfig}

\usepackage{hyperref}
\usepackage{cleveref}

\begin{document}

\title{Memory Profiling and Migration for Heterogeneous Memory Architectures}
 
\author{
Marios Asiminakis, Polydoros Petrakis, and Manolis Marazakis
}

\affiliation{
  \department{Institute of Computer Science}
  \institution{Foundation for Research and Technology - Hellas (FORTH)}
  \city{Heraklion}
  \country{Greece}
}

\begin{abstract}
Heterogeneous memory systems that combine high-bandwidth memory (HBM) with commodity DRAM can accelerate bandwidth-bound HPC workloads, but today’s page placement largely depends on manual tuning or OS heuristics not designed for multi-tier dynamics. We present SHAMBLES, a kernel-integrated framework that profiles application memory behavior at low overhead and migrates data across tiers without requiring application changes. SHAMBLES exposes a policy-agnostic interface and a lightweight user-space runtime with pluggable policies (e.g., recency and frequency based) as well as static placement for controlled studies. A logging mode provides reproducible timelines of allocations and migrations to aid analysis. We implement SHAMBLES on a commodity Linux system with HBM and DDR exposed as NUMA nodes and evaluate it with the HPCG, DGEMM benchmarks and Himeno stencil mini-app. Our design and methodology show how transparent, policy-driven migration can respond to changing access locality and concentrate "hot" data in HBM without developer intervention, offering a practical path to performance portability on tiered memory. Results from HPCG show that we can maintain up to 93.75\% of the \textit{all-in-HBM} baseline performance, while keeping only 40\% of the problem size in the HBM. DGEMM experiments show that SHAMBLES’ dynamic policies sustain up to 99\% of the \textit{all-in-HBM} performance, while keeping only one third of the DGEMM matrix footprint in HBM. 
For Himeno, SHAMBLES shows that fast-tier selection must be workload- and size-aware: with a 50\% fast-tier budget, it can outperform fixed \textit{all-in-HBM/DDR} placements for the L size, while the XL size shifts back toward HBM.

\end{abstract}

\maketitle
\fancyhead[LE]{}
\fancyhead[RO]{}

\section{Introduction}

Modern high-performance computing (HPC) systems and data-intensive applications increasingly rely on heterogeneous memory architectures that combine different memory technologies, such as traditional DRAM and high-bandwidth memory (HBM). These architectures offer significant performance benefits by providing high-speed memory tiers for bandwidth-sensitive workloads while maintaining cost efficiency through larger, slower memory pools. However, efficiently managing data placement across these tiers remains a critical challenge.

%%%
In current systems, memory management is either handled manually by developers or relies on operating system (OS) heuristics. Manual optimization requires extensive profiling and tuning, making it impractical for dynamically changing workloads.
While OS-level mechanisms such as Linux's automatic NUMA balancing (AutoNUMA)~\cite{kernel_numa_balancing} migrate pages automatically, they primarily aim to reduce remote accesses by moving data to the nodes that access it; newer kernels add a memory-tiering mode that places hot pages in fast memory, but its policy is built into the kernel.

Several approaches attempt to address this challenge. Hardware-assisted solutions, such as Intel Memory Mode~\cite{intel_optane_startup_guide},
extend memory capacity by using persistent memory as volatile main memory with DRAM acting as a hardware-managed cache, but they do not provide fine-grained migration control.
Similarly, Heterogeneous Memory Management (HMM) in the Linux kernel~\cite{hmm_linux} enables unified virtual address spaces by integrating device memory (e.g., GPU memory) with the standard memory model,
yet it targets driver-managed device memory rather than providing a placement policy for CPU-attached memory tiers.
Software-based techniques such as the X-Mem allocator~\cite{data-tiering-in-het} place data across tiers but require applications to be modified to tag their allocations.
Profiling tools such as Intel VTune~\cite{intel_vtune_user_guide} and Linux Perf
help analyze memory access patterns but do not themselves place or migrate data; acting on their results requires application modifications or manual tuning.
These limitations highlight the need for a kernel-integrated solution that dynamically adapts memory placement based on runtime behavior without requiring user intervention.
%%%

To bridge this gap, we introduce \textbf{SHAMBLES}, a kernel-integrated memory profiling and migration framework with a lightweight user-space runtime and pluggable policies, designed to automate tier-aware page placement without requiring specialized hardware or changes to application code.
SHAMBLES dynamically monitors memory access patterns and transparently migrates pages across memory tiers based on configurable policies. By integrating both profiling and migration mechanisms within the OS, SHAMBLES enables fine-grained control over data placement while remaining adaptable to different workloads.

SHAMBLES is implemented as a commodity Linux kernel extension with a user-space runtime, and is available as open-source software at \url{https://github.com/CARV-ICS-FORTH/shambles}.

\noindent\textbf{Contributions.} This paper makes the following contributions:
\begin{itemize}
  \item We present SHAMBLES, a Linux kernel--integrated profiling and migration framework that enables transparent, tier-aware data placement for unmodified HPC applications using low-overhead online access sampling and pluggable user-space policies.
  %\item We evaluate SHAMBLES on an Intel Xeon Max HBM+DDR platform and demonstrate that it can reduce fast-tier capacity requirements while maintaining performance, and that its policies are configurable to match the needs of bandwidth-bound and latency-sensitive workloads.
  \item We evaluate SHAMBLES on an Intel Xeon Max HBM+DDR platform and demonstrate that it maintains near-peak performance while reducing HBM
  footprint by 50--66.6\%, and that its policies are configurable to match bandwidth-bound and latency-sensitive workloads.
\end{itemize}

This paper presents the design, implementation, and evaluation of SHAMBLES in an HPC context. We describe its kernel-space and user-space components, highlight its dynamic migration policies, and evaluate it against practical baselines on a real HBM+DDR system. Our evaluation demonstrates that SHAMBLES achieves significant performance improvements by intelligently managing data placement in heterogeneous memory systems.

%%%%%%%%%%%%%%%%%%%%%%%%%%%

\section{Related Work}

Several research efforts have focused on profiling and migrating pages across heterogeneous memory tiers in Linux and Linux-like environments. Most systems target DRAM plus slower capacity tiers such as NVM, CXL-attached memory, or remote DRAM, and typically expose a small number of built-in policies tuned for large-scale datacenter or big-data workloads. By contrast, SHAMBLES provides a pluggable policy framework that targets HBM+DDR nodes in HPC clusters, emphasizes low-overhead sampling of production runs, and is explicitly designed to support both hand-tuned static placements and online, recency-based migration policies on unmodified applications. Although we instantiate and evaluate SHAMBLES on an HBM+DDR platform, the framework itself is technology-agnostic: as long as heterogeneous tiers are exposed as distinct NUMA nodes, the same profiling and policy plug-ins can be applied to DRAM+NVM or CXL-attached memory deployments with minimal changes. However, SHAMBLES primarily targets HPC platforms with CPU-attached tiered memory, rather than CXL-based memory expansion systems. Although this paper evaluates SHAMBLES on an HBM+DDR platform, we have also exercised the framework on a DRAM+NVM system to validate portability across NUMA-exposed tiers (results omitted).

Several systems provide sophisticated multi-tier or CXL-aware tiering mechanisms. \textit{MTM} \cite{ren2024mtm} targets multi-tiered memory hierarchies with a universal migration policy and huge-page-aware management, and reports substantial gains on big-data workloads.
\textit{NeoMem} \cite{zhou2024neomem} relies on hardware/software co-design for CXL devices.
NeoMem adds a device-side profiler, 
while Nomad \cite{nomad2024} introduces transactional page migration with shadow copies to reduce migration cost and memory thrashing under memory pressure.

\textit{TPP} \cite{tpp2024} offers an OS-level placement mechanism for CXL-enabled tiered memory, evaluated with production workloads at Meta and largely merged into Linux v5.18, using lightweight hot/cold classification to close most of the gap to an all-local-memory baseline.
SHAMBLES deliberately stays within commodity CPU and memory hardware, reuses existing kernel mechanisms, and exposes the migration logic as a user-space plug-in interface for experimentation on single-node HBM+DDR platforms rather than as a single in-kernel policy.

Efficient, sampling-based profiling is also central to \textit{DAMON} \cite{damon2020}, a mainline Linux kernel facility that groups pages into regions and periodically samples accesses to reduce profiling overhead. \textit{DAMOS} (DAMON-based Operation Schemes) can then apply user-defined actions on the monitored regions, such as page migration or reclamation, based on access statistics. SHAMBLES follows a similar philosophy of low-overhead sampling and policy-driven actions, but specializes the infrastructure for heterogeneous HBM+DDR platforms and HPC workloads, with an emphasis on exposing profiling data and policy behavior for offline analysis.

A long line of work studies page placement in two-tier memory systems such as DRAM+NVM.
\textit{Thermostat} \cite{thermostat} combines sampling-based tracking, which poisons the page-table entries of sampled pages, with huge-page-aware migration decisions, while the
\textit{Adaptive Page Migration Policy with Huge Pages} \cite{adaptive-huge-pages-tiered} migrates at huge-page granularity based on periodic scans of page-table accessed bits.
\textit{Data Tiering in Heterogeneous Memory Systems} \cite{data-tiering-in-het} proposes X-Mem, a user-space allocator that places application data structures in DRAM or NVM based on offline profiling of their access patterns; applications must be modified to tag their allocations.
Like SHAMBLES, X-Mem builds on \textit{jemalloc} and uses \textit{mbind} to place memory on NUMA-exposed tiers, but it requires source changes and offline profiling.
Closest to our design, \textit{bkmalloc} with the MAT Daemon~\cite{kammerdiener2025bkmalloc} also interposes a custom allocator via \texttt{LD\_PRELOAD} and lets user-level software define object-tiering policies, profiling objects online with perf-based sampling of last-level-cache misses and migrating them from a separate daemon; it was evaluated on DDR4+Optane. SHAMBLES instead samples accesses in the kernel through page-table poisoning, without relying on processor-specific performance events, and targets HBM+DDR nodes.
\textit{Memtis} \cite{memtis-eff-mem-tiering-page-classification}, \textit{Nimble} \cite{nimble1}, and \textit{HeMem} \cite{hemem_2021} 
further improve tiered memory management via dynamic page classification, high-throughput migration, and asynchronous access tracking. 
These systems share SHAMBLES's goal of leveraging software-managed tiering, but they are primarily evaluated on real or emulated NVM, CXL, or disaggregated slow-memory tiers.
HeMem similarly places per-application policy in a user-level library; SHAMBLES instead pairs kernel-level access sampling with a pluggable user-space policy interface aimed at experimentation.
SHAMBLES also decouples profiling from policy and only relies on the generic Linux NUMA abstraction,
so the same infrastructure can be redeployed on DRAM+NVM platforms by mapping the fast and slow NUMA nodes accordingly.

Closer to our focus on HPC, \textit{ecoHMEM} \cite{ecoHMEM} proposes an ecosystem for hybrid DRAM+Optane systems that combines an Extrae- \cite{extrae} and PEBS-based profiler \cite{pebs}, an offline advisor, and a FlexMalloc-based runtime allocator. In a first phase, ecoHMEM profiles object-level behavior and attributes LLC misses to individual allocations; 
an offline “HMem Advisor” then computes a placement of objects in DRAM or 
persistent memory under a DRAM capacity limit,
optionally accounting for each object's bandwidth demand,
 and a second, optimized run enforces this static layout via an interposed allocator.
ecoHMEM outperforms Optane's hardware-managed memory mode on HPC mini-apps when given comparable DRAM,
with smaller or no gains on its two production applications.
SHAMBLES likewise targets HPC workloads and uses an interposed allocator, but performs online chunk-level sampling and dynamic page migration on HBM+DDR nodes, exposing a pluggable set of static and dynamic policies rather than a single offline object-placement advisor.

\textit{Colloid} \cite{colloid} revisits tiered memory management from the perspective of access latency. It observes that existing systems implicitly assume a fixed hierarchy in which the “fast” tier (e.g., local DRAM) always offers lower access latency than the “slow” tier, and shows that under contention this assumption can break down, leading to suboptimal placements. 
Colloid integrates with existing tiering systems such as HeMem, Memtis, and TPP, continuously measures loaded access latencies for each tier, and redistributes hot pages so that observed latencies are balanced across tiers rather than simply filling the nominal fast tier.
This latency-centric view is conceptually related to our Himeno results, where DDR behaves as the preferable tier for a latency-bound stencil on Xeon Max; however, 
Colloid augments existing tiering systems and is evaluated on real-world applications such as graph processing, transaction processing, and key-value caching, whereas SHAMBLES focuses on single-node HBM+DDR HPC nodes and offers a user-space plugin framework that can be reconfigured to treat either HBM or DDR as the effective fast tier. SHAMBLES likewise supports selecting which NUMA nodes constitute the \textit{fast} tier via configuration, but it currently relies on user-provided policies and does not yet adjust this mapping automatically based on online latency measurements as Colloid does.

Other work focuses on large-scale, remote, or virtualized memory. Google's \textit{Software-Defined Far Memory}
\cite{far-memory-warehouse-scale-computers} creates a logical far-memory tier in software by compressing cold pages and combining kernel support with node-level control to meet service level objectives at lower memory cost. 
\textit{TMO} \cite{tmo-memory-offload}, deployed at Meta, offloads cold memory to heterogeneous devices such as compressed memory or SSD-backed tiers, using Linux pressure-stall information to quantify resource pressure (CPU, memory, and I/O) and automatically tune how much memory to offload. \textit{Pond} \cite{pond-cxl} uses CXL to pool DRAM across sockets in cloud platforms and allocates local versus pooled memory per VM to remain close to same-NUMA-node performance. \textit{Efficient Memory Tiering in a Virtual Machine} \cite{prakash2025eff_mem_tier} (GPAC) consolidates scattered hot pages inside VMs so that host tiering mechanisms such as TPP see “densely hot” huge pages.
SHAMBLES is complementary to these efforts: it operates within a single node, does not address pooling or virtualization-aware remapping, and instead focuses on HBM+DDR tiering on bare-metal HPC nodes.

Finally, NUMA balancing mechanisms form an important baseline. 
Linux \textit{AutoNUMA} migrates tasks and data across NUMA nodes to improve locality without application changes, but it can also hurt performance; \textit{Global-State Aware AutoNUMA} \cite{global_autonuma} therefore decides at run time whether to enable it.
SHAMBLES builds on the same principle of
transparent, OS-driven placement, but targets heterogeneous HBM+DDR memory hierarchies and offers a rich set of static and dynamic tiering policies, rather than a single built-in NUMA balancing algorithm. Taken together, these systems demonstrate the effectiveness of software-managed tiering, far-memory offloading, and NUMA balancing for a wide variety of datacenter and big-data workloads. SHAMBLES contributes a complementary point in this design space by (i) targeting commodity HBM+DDR HPC nodes, (ii) providing a pluggable policy framework that spans both static placements and dynamic migration policies, and (iii) coupling low-overhead online profiling with detailed offline analysis to help practitioners understand and tune tiering behavior for memory-intensive scientific workloads. 

Because SHAMBLES only assumes that memory tiers appear as separate NUMA nodes, the framework can be ported to other heterogeneous hierarchies such as DRAM+NVM or CXL-attached memory. However, the prototype and evaluation in
this paper focus on CPU-attached HBM+DDR HPC nodes, where the main challenge is explicit management of scarce high-bandwidth memory under an application-level fast-tier budget.

Prior tiering systems are therefore not directly like-for-like baselines for our evaluation: many are designed around different hardware assumptions, deployment models, workloads, or policy interfaces, such as CXL memory
expansion, DRAM+NVM tiering, remote memory, datacenter workloads, or fixed in-kernel policies. We compare against the relevant Xeon Max baselines: \textit{cache mode}, \textit{HBM-only}, \textit{all-in-HBM}, \textit{all-in-DDR}, and selective static placements. Mainline mechanisms such as \textit{AutoNUMA} and \textit{DAMON}/\textit{DAMOS} remain important conceptual baselines; however, matching SHAMBLES' allocation-aware chunk policies and explicit fast-tier capacity constraints would require additional policy integration and tuning beyond their default configurations.

%%%%%%%%%%%%%%%%%%% end of related work

\section{Design and Implementation Overview}

SHAMBLES consists of three main building blocks:

\begin{itemize}
    \item \textbf{Kernel Patches}: The patches that need to be applied to the mainline Linux kernel, in order for SHAMBLES to work.
    \item \textbf{Modified Allocator}: A modified version of the \textit{jemalloc} library, that loads the user-provided plugin.
    \item \textbf{SHAMBLES plugins}: A set of dynamically loaded libraries that implement the various functionalities.
\end{itemize}

The modular design allows implementing various policies in the plugins, without modifying or recompiling other components.
It also allows porting SHAMBLES to other architectures without modifying the source code of the user space components.
The above components, as well as the interactions between them, are presented in more detail below.

\subsection{Linux Kernel Modifications}

The kernel component of SHAMBLES instruments the virtual-memory subsystem to sample accesses at low overhead, while leaving ordinary mappings unaffected. It reserves a disjoint, high-virtual-address band for sampled buffers and introduces an \textit{mmap} flag that directs profiled allocations into this region, while normal mappings are confined below a separate upper bound; this separation prevents profiling actions from interfering with the rest of the process. Sampling is realized by temporarily "poisoning" the respective Page Global Directory (PGD) entry by clearing its present bit, so that the next access anywhere in the covered range triggers an intentional page fault that serves to capture an access sample. After recording the faulting address, the kernel restores the present bit to resume normal execution. This poisoning is repeated at configurable intervals by a kernel timer.

Operating at upper-level directory granularity amortizes costs across many pages (and aligns with huge pages), avoiding the overhead of per-page toggling. A lightweight kernel–user interface exposes controls and streams sampled addresses, as well as the PC of the instruction that triggered each access, to the user space, via the \textit{debugfs} interface. This information is used by the plugins to implement the desired policy and other functionalities; the kernel remains policy-agnostic and can support both x86 and ARM architectures, with the potential of porting to other architectures.

For both architectures, SHAMBLES currently supports four-level page tables with 4KB base pages and huge pages. We do not currently support five-level page tables on x86, nor 16KB or 64KB base page configurations on ARM. The two ISAs differ in page-table formats and TLB invalidation mechanisms; SHAMBLES relies on existing kernel abstractions to encapsulate most of these ISA-specific details and ease future porting.

\subsection{Modified Allocator}

The user-space allocator is based on \textit{jemalloc} and overrides the default allocator using \texttt{LD\_PRELOAD}. Thus, SHAMBLES is enabled without any modifications or recompilation of the application.
It utilizes the \textit{mmap} flag that we implemented in the kernel, in order to allocate monitored memory addresses. It also dynamically loads the plugin specified by the \texttt{SHAMBLES\_PLUGIN} environment variable, and interfaces with it during each allocator call (allocation, reallocation, free).

\subsection{SHAMBLES plugins}

Most of the SHAMBLES logic is implemented in the plugins. They need to implement the \textit{shambles\_init} function which is called by the custom allocator. This function should initialize the plugin and return a callback function that is called by the allocator when an allocation/reallocation/deallocation event happens. During initialization, the plugin usually spawns a thread that reads the memory access info provided by the kernel and based on that decides if a migration should be triggered. The callback and this thread, also update the counters or the log, if this functionality is enabled.
This is a flexible architecture that allows users to customize memory profiling and migration strategies without modifying the above parts. The plugin mechanism operates at the user-space level, leveraging the modified version of \textit{jemalloc} presented above, to apply different placement and migration policies at run-time.

%%\section{Dynamic Memory Migration Policies}
\section{Implemented Policies}

We make the assumption that one set of NUMA nodes has fast memory, while another set has slow memory. For each policy, several parameters are tunable using environment variables. Plugins operate on allocation objects, which are further split into chunks. We have implemented the following policies:

\begin{itemize}
    \item \textbf{Static}: The allocated memory is placed in a predefined memory type during allocation and is never migrated.
    \item \textbf{LRU}: The most recently accessed chunks are kept in the fast memory and the least recently accessed ones in the slow one.
    \item \textbf{Window}: The placement of the chunks is decided by the frequency of each chunk's accesses during a sliding window of samples.
\end{itemize}

The latter two policies can cause memory migrations during the execution of the application and therefore we call them dynamic policies.
Their aim is to enhance performance by adapting to the application's memory access patterns in real-time. Our architecture ensures that both dynamic migration policies and static placement policies can be deployed seamlessly, allowing for either adaptive or predetermined memory allocation strategies. 

Furthermore, for each policy, there are multiple plugin variants, based on the logging level:  a) \textbf{No Logging}, where no information is stored about allocations, accesses and migrations; b) \textbf{Counters}, where event counts are stored in an \textit{mmap}-ed file and read after execution. It provides useful information with negligible overhead, and Counters mode was used in our evaluation; and c) \textbf{Log}, where a complete, timestamped log is kept for every event. In Log mode SHAMBLES records timestamps, allocation and de-allocation events, virtual addresses, allocation sizes, and every migration with its direction. This mode enables post-hoc timeline reconstruction and correlation with application phases. Log mode increases memory and I/O overhead, and therefore is intended for exploration runs, while production runs typically rely on lightweight counters and statistical summaries (Counters mode).

\subsection{Common Concepts Across Implemented Policies}

SHAMBLES treats a user-defined subset of NUMA nodes as the \emph{fast tier} and the rest as the \emph{capacity (slow) tier}. Policies decide per chunk which tier to use; the runtime enforces this via NUMA bindings. The fast tier is selected with \texttt{SHAMBLES\_FAST\_NODEMASK} and can be capped with \texttt{SHAMBLES\_FAST\_MEM\_SIZE}, so the same binaries/plugins can target HBM, DDR, or any NUMA-exposed memory.
% Placement (and optional page movement) is requested via \textit{mbind} when migrations are enabled.
SHAMBLES tuning is consistent across the plugins. All policies use the following environment variables:
\begin{itemize}
    \item \texttt{SHAMBLES\_SIZE\_THRESHOLD}: The size below which the allocations are not handled by SHAMBLES. 
    \item \texttt{SHAMBLES\_FAST\_NODEMASK}: A bitmap that describes the set of the fast memory nodes.
    \item \texttt{SHAMBLES\_SLOW\_NODEMASK}: A bitmap that describes the set of the slow memory nodes.
\end{itemize}

These variables allow tuning the threshold and the node mask for the specific application as well as the specific NUMA configuration of the machine.
In our experiments, we default to treating HBM NUMA nodes as the fast tier and DDR as the slow tier, unless the workload is latency-sensitive and requires the opposite (e.g. Himeno, Section \ref{sec:himeno}).

Both dynamic policies, also use the following tunables:
\begin{itemize}
    \item \texttt{SHAMBLES\_FAST\_MEM\_SIZE}: The available fast-tier capacity (in bytes). This budget can be set below the physical fast-tier size to intentionally constrain fast memory to a fraction of the application working set, often preserving most of the performance benefits while reducing fast-tier usage.
    \item \texttt{SHAMBLES\_FAST\_MEM\_CHUNKS}: The number of chunks, in which the fast memory is divided.
    \item \texttt{SHAMBLES\_CHUNK\_SIZE}: The size of each chunk (in bytes).
\end{itemize}

Only two of the above three variables need to be set by the user, and the other can be automatically deduced.
The above variables allow the user to explicitly limit the fast memory available to the application. They also control how the allocation objects are split into chunks. Additionally, the user must specify the appropriate CPU and memory affinities, using numactl or a similar method.

In all applicable policies, the initial placement as well as the migrations are performed using the \textit{mbind} system call with the \texttt{MPOL\_BIND} mode and the \texttt{MPOL\_MF\_MOVE} flag, which is available in mainline Linux, to bind the needed virtual address range to a specific set of NUMA memory node.
The time complexity of the processing of each sample is constant, both for logging purposes and for migration decisions for the dynamic policies.
However, the migrations take longer and are more disruptive to application execution as the chunk size increases. The effect is amplified when huge pages are moved.

\subsection{Static Policy}

The \textit{static-fractional} plugin provides fine-grained control over memory placement for specific data allocations. Instead of dynamically migrating pages, it allows users to specify the fraction of each allocation that should reside in fast memory (e.g., HBM) versus slower memory (e.g., DDR). Memory placement is specified as a list of floating-point values representing the desired percentage of an allocation in a memory tier. 

\subsection{Least Recently Used (LRU) Policy}

The LRU policy is designed to leverage temporal locality by ensuring that recently accessed data resides in faster memory tiers. The key components of this policy include:
\begin{itemize}
    \item \textbf{Access Monitoring:} SHAMBLES utilizes page fault counters and address range monitoring to track accesses.
    \item \textbf{Chunk Ranking:} Chunks are ranked based on their last access times. More recently accessed ("hot") chunks are prioritized for placement in fast memory, while less recently accessed ("cold") chunks are relegated to slow memory.
    \item \textbf{Migration Mechanism:} The chunk in which a sample is found to belong is moved to the head of the LRU list. If the chunk is not already in fast memory, it is moved there, while the tail of the LRU list is moved to slow memory.
\end{itemize}

\subsection{Window-Based Policy}

The Window-Based policy focuses on the frequency of memory accesses within a specific window, allowing SHAMBLES to adapt to changing access patterns. This policy operates as follows:

\begin{itemize}
    \item \textbf{Sliding Window Monitoring:} SHAMBLES maintains a sliding window that records the most recent $N$ accesses, that belong to allocations at least as large as the threshold, where $N$ is a configurable parameter (via the \texttt{SHAMBLES\allowbreak\_WINDOW\allowbreak\_SIZE} environment variable).
    \item \textbf{Frequency Analysis:} Within this window, the plugin maintains the access frequency for each chunk. Pages with higher access counts  are allocated to faster memory tiers, while those with lower frequencies are migrated to slower memory.
    \item \textbf{Adaptive Migration:} When a sample is received, SHAMBLES reassesses the access frequencies and adjusts page placements accordingly. This dynamic adjustment ensures that memory allocation aligns with the current working-set of the application, optimizing performance.
\end{itemize}

\subsection{Example runtime configuration}
\label{sec:shambles-example-config}

%As discussed above, SHAMBLES is configured entirely via environment variables. 
The following example shows how to configure a 
\textit{window(16)-chunks:4} policy scenario, via environment variables:

\begin{Verbatim}[fontsize=\small]
export LD_PRELOAD=./jemalloc/libjemalloc.so
export SHAMBLES_PLUGIN=./plugins/window.so
export SHAMBLES_FAST_NODEMASK=1
export SHAMBLES_SLOW_NODEMASK=2
export SHAMBLES_SIZE_THRESHOLD=1048576 # 1 MiB
export SHAMBLES_FAST_MEM_CHUNKS=4
export SHAMBLES_WINDOW_SIZE=16 # 16 samples window
export SHAMBLES_FAST_MEM_SIZE=12582912 # 12 MiB
\end{Verbatim}

\section{Evaluation Methodology and Results}

\subsection{Tested system: Intel Xeon CPU Max 9468}

We evaluate SHAMBLES on a dual-socket Intel Xeon CPU Max 9468 server (48 cores, 8 DDR5 channels, and 4 HBM2e stacks per socket). The system exposes the on-package HBM and off-package DDR5 tiers as distinct NUMA nodes, enabling SHAMBLES to observe locality and migrate data transparently between them. We focus our quantitative evaluation on Xeon Max, but we have also exercised SHAMBLES on an Intel Xeon Gold 5318Y DRAM+Optane (NVM) platform to confirm functional portability across different tier pairs.
The system runs a 64-bit Linux kernel 6.6.3 incorporating our SHAMBLES patch. CPU frequency is fixed at the base 2.1 GHz, and Turbo-boost is disabled 
to eliminate run-to-run variability from dynamic per-core frequency scaling. Each CPU core has 2 AVX-512 FMA units, yielding a theoretical peak per core of 67.2 Double Precision (DP) GF/s.

On this platform we enable Sub-NUMA Clustering (SNC) in SNC4 mode, which partitions each Xeon Max 9468 socket into four sub-NUMA clusters. Each SNC corresponds to a single compute tile and groups 12 physical cores with its local slice of the last-level cache, a pair of DDR5 channels, and one 16\,GiB HBM2e stack. Importantly, within each SNC the operating system exposes the DDR and HBM tiers as \emph{distinct} NUMA nodes: the DDR NUMA node is the \emph{compute} node hosting the tile's CPU cores, while the HBM NUMA node is \emph{memory-only}. This separation enables SHAMBLES to target each tier explicitly. This configuration reflects modern HPC nodes with tiered memory, providing a testbed for evaluating profiling fidelity and page migration policies under SHAMBLES. The DDR5 and HBM latency and bandwidth for Intel Xeon Max (measured using Intel MLC utility \cite{intel_mlc}) are shown in Table \ref{tab:xeon_lat_bw}.
For our experiments we pin processes to NUMA nodes (SNCs) to respect locality and minimize cross-NUMA interference.

\begin{table}[h]
\centering
\small
\begin{tabular}{cccc}
\hline
\textbf{Socket} & \textbf{SNCs} & \textbf{DDR NUMA} & \textbf{HBM NUMA} \\
\hline
0 & SNC0--SNC3 & 0--3  (12c, 32 GiB each) & 8--11  (16 GiB each) \\
1 & SNC4--SNC7 & 4--7  (12c, 32 GiB each) & 12--15 (16 GiB each) \\
\hline
\end{tabular}
\caption{Intel Xeon Max 9468 NUMA layout in SNC4 mode. Each SNC tile
contains 12 cores (12c), one compute-local DDR NUMA node, and one
memory-only HBM NUMA node.}
\label{tab:xeon_numa_topology}
\end{table}

\begin{table}[h]
\centering
\begin{tabular}{lccc}
\hline
\multicolumn{4}{c}{\textbf{Avg. Latency (ns)}} \\
\hline
\textbf{Tier} & \textbf{Intra-SNC} & \textbf{Intra-socket} & \textbf{Inter-socket} \\
\hline
DDR5 & 100.6 & 116.5 & 238.3 \\
HBM  & 123.7 & 140.0 & 242.8 \\
\hline
\multicolumn{4}{c}{\textbf{Bandwidth (GB/s)}} \\
\hline
\textbf{Tier} & \textbf{Intra-SNC} & \textbf{Intra-socket} & \textbf{Inter-socket} \\
\hline
DDR5 & 60.1  & 60.1  & 53.6  \\
HBM  & 189.6 & 119.6 & 35.4 \\
\hline
\end{tabular}
\caption{Intel Xeon Max 9468 memory latency and bandwidth.}
\label{tab:xeon_lat_bw}
\end{table}

\subsection{Performance results - HPCG}

The High Performance Conjugate Gradients (HPCG) benchmark \cite{hpcg_ref} complements HPL/LINPACK \cite{dongarra2003linpack} by stressing sparse matrix--vector products, vector updates, multigrid smoothing, and global reductions rather than peak FLOP throughput. These kernels are limited mainly by memory bandwidth, latency, and communication, making HPCG a representative memory-intensive workload for evaluating SHAMBLES' dynamic tiering policies.

For our experiments we use the default HPCG problem size $nx=104$, $ny=104$, $nz=104$, which means a total size of 579.28 MiB per MPI Rank. We keep the default sizes of HPCG, with the 27-point stencil and  $local\_int\_size= 4$ bytes and $global\_int\_size$ = 8 bytes. The three main data structures of HPCG are the \textbf{CSR Local Indices Data} (\textit{mtxIndL[0]}), \textbf{CSR Matrix Values Data} (\textit{matrixValues[0]}) and \textbf{CSR Global Indices Data} (\textit{mtxIndG[0]}). We configure the SHAMBLES threshold to 32 MiB, so that all small allocations are handled by the OS, while SHAMBLES is responsible for the monitoring and migration of the three largest allocations. Note that, SHAMBLES also monitors the memory accesses of smaller allocations (below threshold), but does not migrate any of these.  

As a reference, we have run HPCG in \textit{cache} mode and \textit{HBM-only} modes, along with two static placement scenarios. The Intel Xeon Max can be configured so that the HBM can act as an additional Last Level Cache (LLC) L4; we call this \textit{cache} mode. The second reference run we do is the \textit{HBM-only} mode, where we use \textit{numactl}, to place all the allocations of HPCG in the HBM. In the Intel Max CPU the NUMA distance for the DDR (10) is closer than the distance to the HBM (13). Therefore, anything not placed explicitly in the HBM, is placed to the DDR by the OS. The two static scenarios that we present are a) \textit{mtxIndL \& matrixValues in HBM}, and b) \textit{mtxIndL \&mtxIndG in the HBM}. The selected static combinations place ~60\% of the problem size in the HBM ( (115.86 + 231.71) / 579.28 MiB per rank). In comparison, for our dynamic policies (LRU, window), we chose to place only 40\% of the problem size in the HBM, saving a very significant portion of the HBM and trying to maintain performance. We did that by setting \texttt{FAST\_MEM\_SIZE} equal to 231.75 MiB per rank. We also explored several other static placement combinations (e.g., \textit{mtxIndL} and half of \textit{matrixValues} in HBM); however, these results are omitted due to space constraints.

For our tests we have run with two Sampling Rates (SR) for the dynamic policies: 10 Hz, and 100 Hz; the corresponding sampling overhead is quantified in Section~\ref{sampling_section}. For the window policy we are testing with two window sizes: 16 and 128. Regarding the number of \textit{chunks}, we have tested with $N$ = 2, 8, 32, and 64. chunks=$N$, means that \texttt{FAST\_MEM\_SIZE} is split into $N$ allocations, and each of these chunks can be migrated independently from the others. Note that, all migrations are on the level of chunk, not page size.

\begin{figure}[t]
    \centering
    \includegraphics[width=\linewidth]{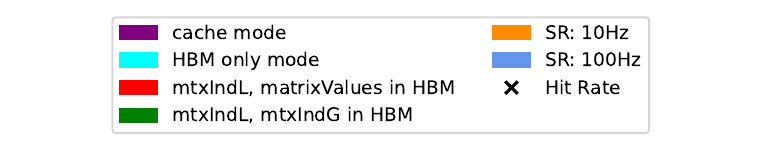}
    \vspace{0.1ex}
    \includegraphics[width=\linewidth]{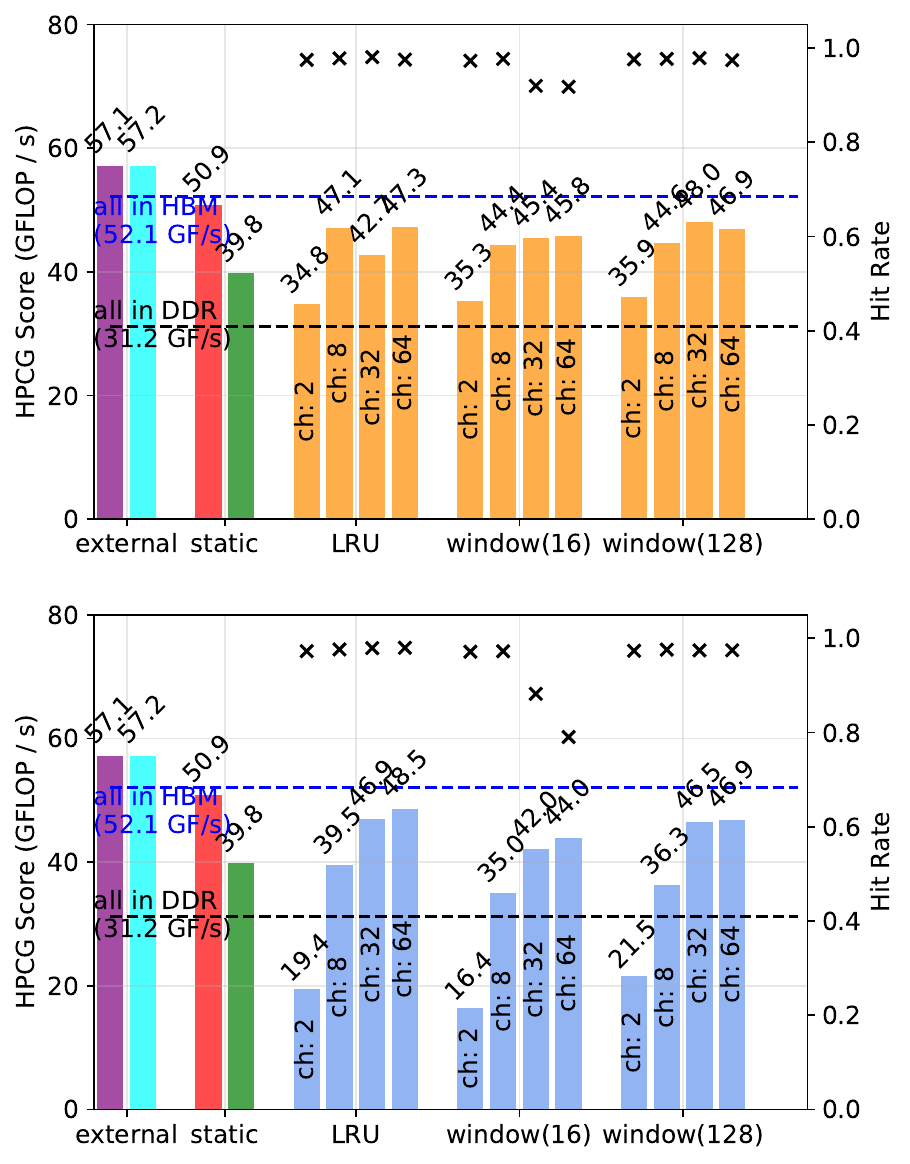}
    \caption{HPCG score for 48 ranks at 10 Hz and 100 Hz Sampling Rates.}
    \label{fig:hpcg48}
\end{figure}

\begin{figure}[t]
    \centering
    %\includegraphics[width=\linewidth]{figures/hpcg/hpcg_legend.pdf}
    %\vspace{0.5ex}
    \includegraphics[width=\linewidth]{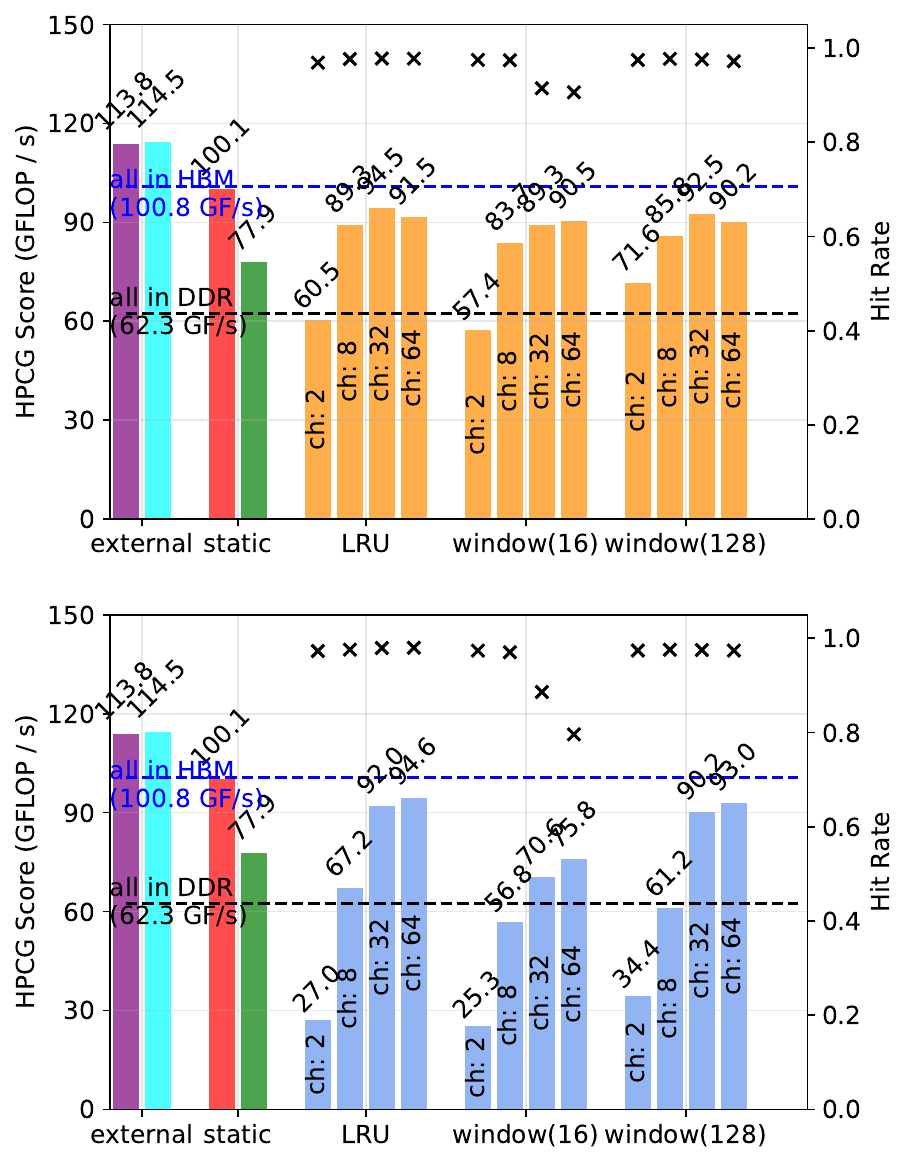}
    \caption{HPCG score for 96 ranks at 10 Hz and 100 Hz Sampling Rates. Legend follows Fig.~\ref{fig:hpcg48}.}
    \label{fig:hpcg96}
\end{figure}

On the following HPCG charts (Fig. \ref{fig:hpcg48} to \ref{fig:hpcg96}), we depict the migration cost for dynamic policies, expressed in GiB, below each chart. The migration cost is summed for all Ranks.   Additionally, we depict the Hit Rate for the 3 large allocations (Large-Allocation-Hits / Large-Allocation-Accesses) on the secondary axis. Finally we depict two dashed lines: \textit{all-in-HBM} and \textit{all-in-DDR}, which means that the three largest allocations are all statically placed on either the HBM, or the DDR (using numactl), while the smaller allocations (below threshold) are placed in the DDR by the OS. Therefore, you will notice a performance difference between the \textit{all-in-HBM} and the static \textit{HBM-only} modes, because \textit{HBM-only} also places the small allocations in the HBM.

When using one socket (48 ranks, 4 SNCs) (see Fig. \ref{fig:hpcg48}), the \textit{cache} mode reaches 57.1 GF/s and the \textit{HBM-only} 57.2 GF/s. So we can conclude that the \textit{cache} mode performs as well as the \textit{HBM-only} mode. The \textit{all-in-HBM} scenario reaches 52.1 GF/s, and we consider this our best baseline scenario, as all the large data allocations are 100\% placed in the HBM, while the smaller allocations are all kept in the DDR. The static scenario of \textit{mtxIndL \& MatrixValues in HBM} reaches 50.9 GF/s (while 60\% of the problem size is in the HBM). The dynamic policies on the other hand, can reach up to 48 GF/s (\textit{window(128)-chunks:32, 10 Hz}), which means \textbf{92.13\% of the best baseline \textit{all-in-HBM} scenario, while keeping only 40\% of the problem size in the HBM}. When using a small number of chunks (2), the performance is degraded for all the dynamic migration policies (<36 GF/s for 10 Hz, and <= 27 GF/s for the 100 Hz cases). However, when we increase the number of chunks (8, 32 or 64), for 10 Hz, \textit{LRU-chunks:64} reaches 47.3 GF/s (\textbf{90.78\% of the best baseline} \textit{all-in-HBM} scenario) and \textit{window(128)-chunks:32} reaches 48 GF/s as mentioned before. For the increased SR of 100 Hz, \textit{window(16)} does not perform so well, as hit rate drops below 80\% and migration cost increases (23.88$\times$ more GiB migrated for the 64 chunks case). Using 100 Hz improves only the \textit{LRU-chunks:64} policy (48.5 GF/s). 

For two sockets (96 ranks, 8 SNCs), shown in Fig. \ref{fig:hpcg96}, the \textit{all-in-HBM} scenario reaches 100.8 GF/s and the \textit{HBM-only} 114.5 GF/s. Using 10 Hz, instead of 100 Hz proves to be the best choice for almost all dynamic scenarios, except \textit{LRU-chunks:64} and \textit{window(128)-chunks:64} cases, which perform similarly to the 10 Hz setups. Obviously, increasing the sampling rate by 10$\times$, can lead to increasing the migrations by 10$\times$. However, we see the migration size increased by 17.66$\times$ (364 GiB from 20.6 GiB), in the \textit{window(16)-chunks:64} case, paired with a large reduction in the hit rate (<80\%). For the 10 Hz, the best-performing dynamic policy setups, shown in Fig. \ref{fig:hpcg96}-top, are a) the \textit{LRU-chunks:32}, which reaches 94.5 GF/s (93.75\% of best baseline), b) the \textit{window(16)-chunks:64}, which reaches 90.5 GF/s (89.78\% of best baseline) and c) the \textit{window(128)-chunks:32} reaches 92.5 GF/s (91.76\% of best baseline).

When using only two chunks, the dynamic policies performance stays very low; even below the \textit{all-in-DDR} mode for most cases and independently of the SR. We believe this is because of the huge increase in the migration overhead, as the migrated data size ranges from 56.4 to 193 GiB for the 10 Hz SR, and 539.1 - 1526.3 GiB for the 100 Hz SR. In comparison, the top performing \textit{LRU-chunks:32, 10Hz}, and \textit{LRU-chunks:64, 10Hz} only migrate 10 GiB and 5GiB respectively, while achieving 94.5 and 91.5 GF/s.

% HPCG conclusion
Overall, SHAMBLES recovers most of the \textit{all-in-HBM} performance for HPCG while using only 40\% of the problem size in HBM, reaching 92.13\% at 48 ranks and 93.75\% at 96 ranks. The best configurations use sufficiently fine chunks and low sampling rates; coarse chunks and 100~Hz sampling increase migration traffic and often reduce performance.

\subsection{Performance results - DGEMM}

For DGEMM, we use the ACES multi-threaded DGEMM benchmark \cite{mt_dgemm}, which calls Intel MKL \texttt{cblas\_dgemm} \cite{intel_onemkl} to compute \( C \leftarrow \alpha AB + \beta C \). DGEMM has high arithmetic intensity and regular memory accesses, making it a compute-bound workload that complements the more memory-sensitive kernels in our evaluation. We run an $8192\times8192$ problem on one SNC domain with 12 OpenMP threads. The three matrices occupy 1536 MiB in total, while SHAMBLES is limited to a 512 MiB fast-tier budget, i.e., only one third of the matrix footprint can reside in HBM. For space, we summarize the DGEMM results in text.

The omitted DGEMM results show that SHAMBLES preserves near-peak performance despite this restricted HBM budget. At 10 Hz SR, the best dynamic policy reaches 763.6 GF/s, or 99.0\% of the \textit{all-in-HBM} baseline, while migrating only 1 GiB over 80 iterations. This is very close to the \textit{all-in-HBM} result of 770.9 GF/s, even though SHAMBLES uses only one third of the HBM capacity required by full HBM placement. Therefore, for DGEMM, SHAMBLES maintains near-\textit{all-in-HBM} performance with substantially reduced HBM usage.

\subsection{Performance results - Himeno}
\label{sec:himeno}

The Himeno benchmark~\cite{HimenoBMT} is a 3-D Jacobi stencil mini-app that solves the pressure Poisson equation and is commonly used to evaluate memory-system behavior. We first study the L problem size ($256\times256\times512$ cells), which has a 1792 MiB working set, and then compare against XL ($512\times512\times1024$), whose main array footprint is about 14 GiB. For both L and XL SHAMBLES \texttt{FAST\_MEM\_SIZE} is configured to 50\% of the working set size (e.g. 896 MiB for L).

The allocation layout follows the conventional structure: solution grid \textit{p}, workspaces \textit{wrk1}/\textit{wrk2}, coefficient arrays
\textit{a}, \textit{b}, \textit{c}, and boundary mask \textit{bnd}, all allocated as contiguous linearized 3-D arrays. This mix of grids,
coefficients, boundary data, and temporary workspaces creates heterogeneous access behavior across arrays, making Himeno well suited for evaluating
selective page placement across HBM and DDR.

\begin{figure}[t]
    \centering
    \includegraphics[width=1\linewidth]{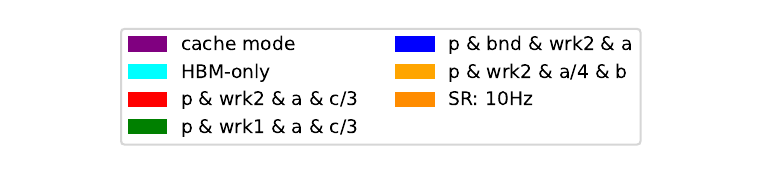}
    %old_label: himeno_100hz
    \vspace{0.1ex}    
    \includegraphics[width=1\linewidth]{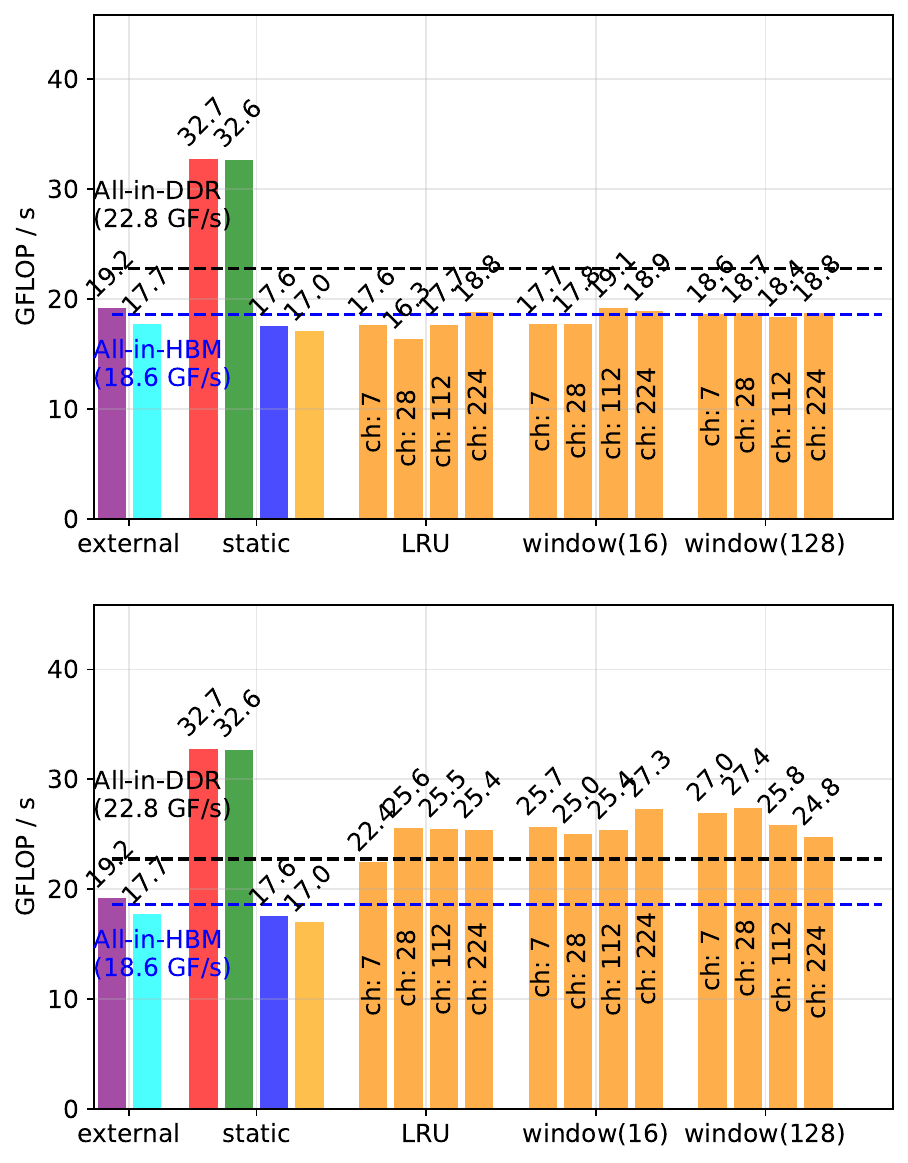}
    %old_label: himeno_100hz_reversed
    \caption{Himeno Performance Large (GF/s), 12 threads, 10Hz, \emph{Top}: HBM is considered the fast tier, \emph{Bottom}: DDR is considered the fast tier}   
    \label{fig:himeno_L_top_bottom}
\end{figure}

As shown in the top panel of Fig. \ref{fig:himeno_L_top_bottom}, for Himeno the \textit{all-in-DDR} can achieve a significantly better performance (22.8 GF/s), versus the \textit{all-in-HBM} scenario, which only achieves 18.6 GF/s. This indicates that, for the L problem size, Himeno does not benefit uniformly from placing all data in HBM: on this platform, the lower-latency DDR path can outperform the higher-bandwidth HBM path for this stencil workload. We omit the hit-rate axis from the Himeno figures for readability and clarity, since the performance trends are primarily explained by tier selection and placement sensitivity rather than by a monotonic hit-rate/performance relationship.

We also evaluated 148 static placements of the largest Himeno arrays. The best layouts reach up to 32.7 GF/s, outperforming the \textit{all-in-DDR} configuration by 43\%, while the worst layouts drop to 17.0 GF/s. This shows that selective HBM use can help, but only with the right placement.

With HBM treated as the fast tier, most SHAMBLES dynamic policies remain close to the \textit{all-in-HBM} baseline, despite using about 50\% of the problem size in HBM. This behavior suggests that, for the L problem size, HBM-fast migration is limited by the tier-choice mismatch and by migration overhead.

After observing that \textit{all-in-DDR} outperforms \textit{all-in-HBM}, we reconfigured SHAMBLES so that DDR is treated as the fast tier by changing
only the relative NUMA node masks.

At 10 Hz SR (bottom panel of Fig.~\ref{fig:himeno_L_top_bottom}), LRU and the best \textit{window} configuration reach 25.6 and 27.4 GF/s
(+12.28\% and +20.17\% vs. \textit{all-in-DDR}); at 100 Hz SR (not shown), they reach 28.2 and 29.0 GF/s (+23.68\% and +27.19\%). This confirms that SHAMBLES benefits from workload-aware tier selection rather than assuming HBM is always the fast tier.
%% XL discussion

For the XL problem size, the tier preference changes compared to the L case. Here, \textit{all-in-HBM} outperforms \textit{all-in-DDR}, reaching 24.3 GF/s versus 20.7 GF/s. This shows that, as the working set grows, Himeno becomes more able to benefit from HBM bandwidth. In contrast, cache mode drops to 11.8 GF/s, which is expected because the XL working set exceeds the HBM capacity available within one SNC domain.

\begin{figure}[t]
    \centering
    \includegraphics[width=1\linewidth]{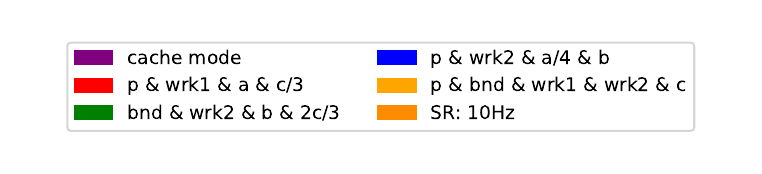}
    \vspace{0.1ex}    
    \includegraphics[width=1\linewidth]{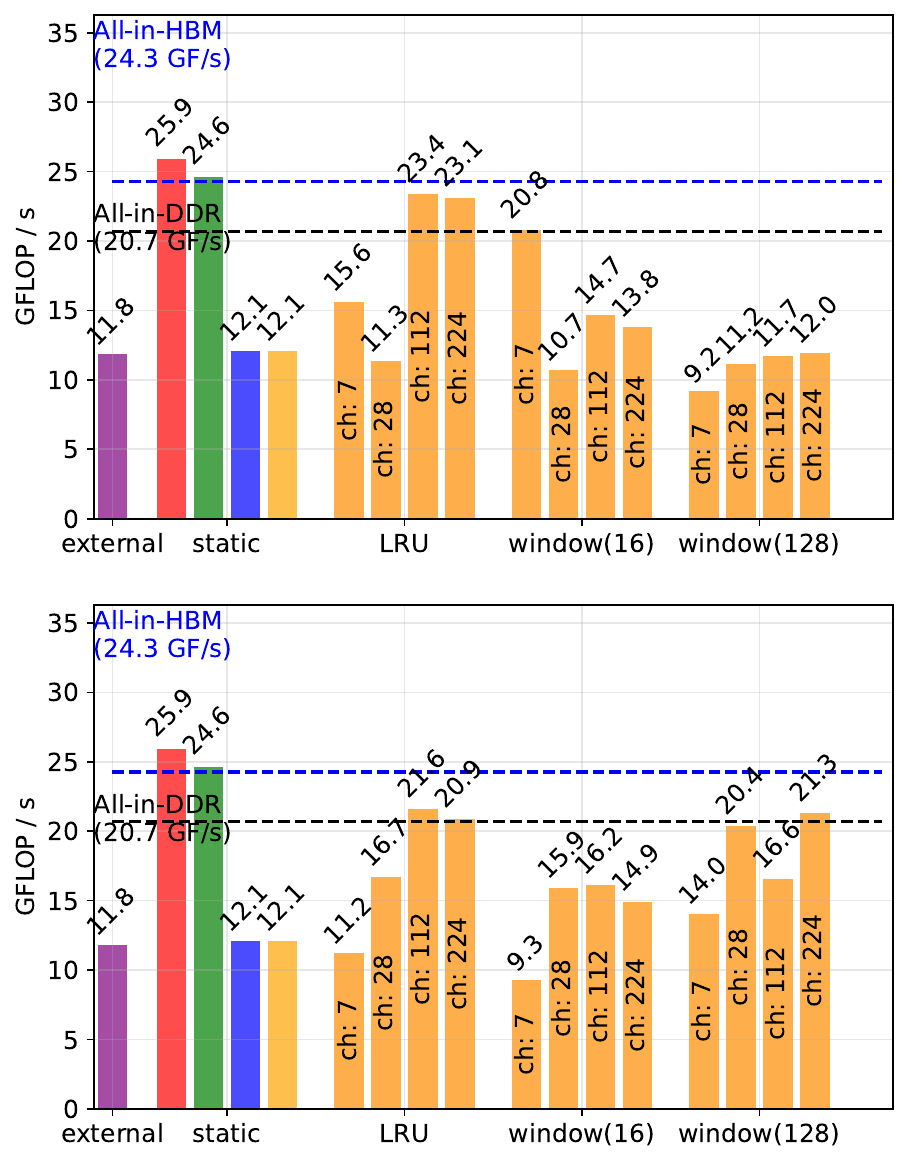}
    \caption{Himeno Performance XL (GF/s), 12 threads, 10 Hz, Top: HBM is considered the fast tier, Bottom: DDR is considered the fast tier}
    \label{fig:himeno_XL_10Hz_top_bottom}
\end{figure}

Fig.~\ref{fig:himeno_XL_10Hz_top_bottom} compares SHAMBLES at 10 Hz when HBM or DDR is treated as the fast tier. For XL, HBM-fast placement is preferable: the best dynamic configuration, \textit{LRU-chunks:112}, reaches 23.4 GF/s, or 96.3\% of the \textit{all-in-HBM} baseline. When DDR is treated as the fast tier, the best dynamic configuration reaches only 21.6 GF/s. Static selective placement can still outperform the \textit{all-in-HBM} and \textit{all-in-DDR} baselines, with the best static layout reaching 25.9 GF/s. Overall, the XL results confirm that the best tier choice is workload- and size-dependent: the L problem benefits from DDR-as-fast placement, while the larger XL problem shifts back toward HBM-as-fast placement.

Overall, Himeno shows that memory-tier ordering is not fixed even for the same application. The L problem benefits from DDR-as-fast placement, whereas XL shifts back toward HBM-as-fast placement, confirming that tiering policies must account for problem size and latency--bandwidth trade-offs rather than assuming that HBM is always the fast tier. The XL results also expose a limitation of cache mode, whose performance drops sharply once the working set exceeds local HBM capacity.

% HIMENO LARGE

%%%%%% HIMENO XL startrs here %%%%%%

\subsection{SHAMBLES Sampling overhead}
\label{sampling_section}

Table~\ref{tab:dgemm_shambles_sampling_overhead} shows SHAMBLES sampling overhead for DGEMM (12 threads). Up to 1{,}000~Hz the overhead stays below 0.6\%; at our default 10--100~Hz rates it is at most 0.08\%, dropping to zero for applications that do not preload SHAMBLES. Higher rates (e.g., 1{,}000~Hz) are useful only in full-system simulators, where workloads run for seconds of simulated time and low SRs yield too few samples. 
For reference, DAMON's authors report that it slows target workloads down by 1.16\% \cite{damon_patch2021}.

\begin{table}[h]
\centering
\small
\begin{tabular}{rrrr}
\hline
\textbf{SR (Hz)} & \textbf{Samples} & \textbf{GF / s} & \textbf{Slowdown vs.\ 0 Hz (\%)} \\
\hline
0      & 0       & 765.23 & 0.00  \\
10     & 115     & 765.64 & -0.05 \\
100    & 1,154    & 764.59 & 0.08  \\
1,000   & 11,595   & 760.94 & 0.56  \\
10,000  & 120,211  & 734.73 & 3.99  \\
100,000 & 1,709,938 & 511.12 & 33.21 \\
\hline
\end{tabular}
\caption{DGEMM performance (12 threads) under different SHAMBLES SRs. Performance drop is computed relative to the 0~Hz baseline.}
\label{tab:dgemm_shambles_sampling_overhead}
\end{table}

\section {Future extensions}

Our current SHAMBLES prototype selects hot pages solely based on sampled page-fault activity. However, it does not receive any feedback from the application or the system about the impact of its migration decisions. For example, SHAMBLES may migrate a set of pages and the application may subsequently reach near-peak GF/s, but SHAMBLES itself remains unaware that this decision was beneficial. An interesting direction for future work is to close this loop by feeding system-level performance signals back into SHAMBLES. One concrete option would be to incorporate hardware performance counters —such as instructions per cycle (IPC), memory stall cycles, or bandwidth utilization— as additional inputs to guide and refine migration policies.
For example, the MAT Daemon used with bkmalloc~\cite{kammerdiener2025bkmalloc} can trigger object migrations based on changes in last-level-cache misses per instruction, although this signal detects program phase changes rather than measuring the benefit of individual migrations.

Another useful, but challenging to implement, feature is automating the tuning of SHAMBLES for the specific platform and application. This includes optimizing the policy selection, the SR and the environment variables. This might require the implementation of micro-benchmarks that extract machine dependent characteristics and on-the-fly profiling of the application being run using performance counters. Specifically for the memory tier selection, the implementation of advanced policies that model low-latency and high-bandwidth tiers, replacing the current "fast" and "slow" tiers, can handle effectively the issue detected when running Himeno.

Finally, the user space components can be adapted to support alternative sampling methods -specifically, hardware assisted methods, like Intel PEBS or Arm SPE \cite{arm_spe_methodology_2023}, or a sampler based on DAMON.
Such alternatives however, have the downsides of lack of portability or a higher performance overhead.

\section{Conclusions}

Heterogeneous memory systems promise both performance and capacity by combining high-bandwidth memory with conventional DRAM, but they also expose a difficult data-placement problem that static policies and generic NUMA heuristics handle poorly. This work introduced SHAMBLES, a kernel-integrated memory profiling and migration framework that operates transparently to applications, requires no specialized hardware, and exposes a policy-agnostic interface with pluggable user-space policies. By combining lightweight page-fault sampling in the kernel with a pre-loadable \textit{jemalloc}-based allocator, SHAMBLES can track hot regions at chunk granularity, migrate them across tiers, and emit detailed logs for post-hoc analysis.

Our evaluation on a dual-socket Intel Xeon CPU Max 9468 platform shows that SHAMBLES can substantially reduce HBM footprint while retaining most of the performance of idealized baselines. For the memory-bound HPCG benchmark, tuned LRU and window-based policies maintain up to 93.75\% of the \textit{all-in-HBM} performance while placing only 40\% of the problem size in HBM, compared to $\approx$60\% for the best static placements. For the compute-bound DGEMM workload, SHAMBLES’ dynamic policies sustain up to 99\% of the \textit{all-in-HBM} baseline while allowing only one third of the matrix footprint to reside in HBM at any given time, effectively saving about two thirds of the HBM capacity that an \textit{all-in-HBM} configuration would require.

Himeno, especially the L problem size, highlights both the pitfalls and the potential of memory tiering on asymmetric platforms. On our system, \textit{all-in-DDR} is faster than \textit{all-in-HBM}; naive recency-based policies that treat HBM as universally “fast” therefore hover around the \textit{all-in-HBM} baseline and pay unnecessary migration overhead. After reconfiguring SHAMBLES so that DDR is treated as the fast tier, the same dynamic policies deliver 12–27\% speedups over the \textit{all-in-DDR} baseline and close much of the gap to the best hand-tuned static layouts.
On the other hand, the XL size, demonstrates that, besides the application, the problem size affects the optimal placement.
These results underscore that effective tiering must respect both latency–bandwidth trade-offs and platform asymmetry, rather than assuming a single linear speed hierarchy.
Finally, comparing the cache mode performance between the two problem sizes, confirms that the cache mode receives a severe performance hit when the memory footprint surpasses the available HBM.

Overall, SHAMBLES demonstrates that OS-integrated, policy-driven migration can provide a practical path to performance portability on tiered memory systems. It decouples profiling and migration mechanisms from policy logic, enables rapid experimentation with dynamic and static placements, and surfaces rich traces for analysis. Future extensions include using hardware performance counters in policy decisions and expanding the policy space to better handle latency-sensitive codes and multi-tenant environments.

\begin{acks}
We thankfully acknowledge support for this research from the European High Performance Computing Joint Undertaking (EuroHPC JU) under Framework Partnership Agreement No 800928 (European Processor Initiative) and Specific Grant Agreement No 101036168 (EPI-SGA2). The EuroHPC JU receives support from the European Union’s Horizon 2020 research and innovation programme and from Croatia, France, Germany, Greece, Italy, Netherlands, Portugal, Spain, Sweden, and Switzerland. National contributions from the involved state members (including the Greek General Secretariat for Research and Innovation) match the EuroHPC funding. The research reported in this paper was also supported in part by the 
European Health and Digital Executive Agency (HaDEA) under Grant Agreement No 101092993 (RISER).
\end{acks}

\bibliographystyle{ACM-Reference-Format}
\bibliography{references}

\end{document}